\documentclass[]{spie}  

\usepackage{subfigure}
\usepackage{bbm}
\usepackage{amsmath,amsfonts,amssymb}
\usepackage{graphicx}
\usepackage{hyperref}
\hypersetup{colorlinks,allcolors=blue}

\newcommand{\argmax}{\operatornamewithlimits{arg\,max}}

\title{Comparing Bayesian Models for Organ Contouring in Head and Neck Radiotherapy}

\author[a]{Prerak P Mody}
\author[b]{Nicolas Chaves-de-Plaza}
\author[b]{Klaus Hildebrandt}
\author[c]{Ren\'e van Egmond}
\author[c]{Huib de Ridder}
\author[a]{Marius Staring}
\affil[a]{Department of Radiology, Leiden University Medical Center, Leiden, The Netherlands}
\affil[b]{Computer Graphics and Visualization Lab, TU Delft, Delft, The Netherlands}
\affil[c]{Industrial Design Engineering, TU Delft, Delft, The Netherlands}
\authorinfo{Further author information: (Send correspondence to P.P.M.)\\P.P.M.: E-mail: P.P.Mody@lumc.nl}

\begin{document} 
\maketitle

\begin{abstract}
Deep learning models for organ contouring in radiotherapy are poised for clinical usage, but currently, there exist few tools for automated quality assessment (QA) of the predicted contours. Bayesian models and their associated uncertainty, can potentially automate the process of detecting inaccurate predictions. We investigate two Bayesian models for auto-contouring, DropOut and FlipOut, using a quantitative measure -- expected calibration error (ECE) and a qualitative measure -- region-based accuracy-vs-uncertainty (R-AvU) graphs. It is well understood that a model should have low ECE to be considered trustworthy. However, in a QA context, a model should also have high uncertainty in inaccurate regions and low uncertainty in accurate regions. Such behaviour could direct visual attention of expert users to potentially inaccurate regions, leading to a speed-up in the QA process. Using R-AvU graphs, we qualitatively compare the behaviour of different models in accurate and inaccurate regions. Experiments are conducted on the MICCAI2015 Head and Neck Segmentation Challenge and on the DeepMindTCIA CT dataset using three models: DropOut-DICE, Dropout-CE (Cross Entropy) and FlipOut-CE. Quantitative results show that DropOut-DICE has the highest ECE, while Dropout-CE and FlipOut-CE have the lowest ECE. To better understand the difference between DropOut-CE and FlipOut-CE, we use the R-AvU graph which shows that FlipOut-CE has better uncertainty coverage in inaccurate regions than DropOut-CE. Such a combination of quantitative and qualitative metrics explores a new approach that helps to select which model can be deployed as a QA tool in clinical settings.
\end{abstract}


\section{INTRODUCTION}
\label{sec:intro} 

Radiotherapy is an important cancer treatment option due to its ability to treat cancerous tissue while simultaneously sparing healthy tissue \cite{global_yap2016}. During treatment planning there is a requirement to acquire diagnostic 3D images like CT, MR and PET scans and contour the healthy tissue or organs at risk (OAR) as well as tumorous tissue. This contouring task is time-consuming and is also subject to inter- and intra-annotator disagreement \cite{observer_3dvariability_brouwer2012, rt_variation_hn_van2019}. As deep learning models have made great progress in this field \cite{anatomynet_zhu2019, focusnet_gao2019, nas_guo2020, focusnetv2_gao2021, novel2D3D_chen2021} they are widely being considered as an automated technique to speed up and standardize the contouring process \cite{dl_assessment_brouwer2020, dl_assessment_van2020}. However, to deploy such models in a clinical setting, a manual quality assessment (QA) of predicted contours needs to be performed before they can be used for radiation dosage calculation, which again, introduces a delay. This work investigates the potential usage of uncertainty heatmaps produced by Bayesian deep learning models to help speed up the manual QA process for OARs, by directing human attention to inaccurately segmented regions.

Organ contours are extracted by classifying the 3D voxels of a scan into different categories. It is well accepted that for a predictive classification model to be trusted, it should be calibrated. This means that its output confidence (i.e. probability value) should correspond to the likelihood of being accurate. In other words, in a calibrated model, voxels predicted to belong to an OAR with probability $p$, should have an accuracy equal to $p$. It has been previously shown that well-calibrated model confidences also produce uncertainty measures that correspond to inaccurate regions \cite{calibrated_sander2019, calibrated_mehrtash2020}. Such a property may be useful in a radiotherapy QA context to direct visual attention of clinicians to inaccurate regions. Thus, this work further investigates this claim, for the purpose of choosing a model for clinical deployment, by analysing two deep Bayesian models - DropOut \cite{dropout_segnet_badrinarayanan2017} and FlipOut \cite{flipout_labonte2019we}. Bayesian models were chosen as they offer a principled approach to capture uncertainty. We use a combination of a commonly used quantitative metric for model confidence calibration - expected calibration error (ECE) \cite{ece_guo2017} and propose a new qualitative metric for uncertainty calibration - region-based accuracy-vs-uncertainty (R-AvU) graphs. Motivated by the observation that some models may provide us with similar ECE values, we use the R-AvU graphs to understand the differences in their uncertainty behavior. Previous uncertainty evaluation metrics like AvU \cite{evaluating_mukhoti2018} provide a single scalar value by performing an analysis on the accuracy and uncertainty of each voxel in a scan. To achieve a perfect AvU score, a model must have only accurate and certain or inaccurate and uncertain voxels, i.e. perfectly calibrated uncertainty. We believe this metric has the right motivation, but its formulation may not be sufficient from a QA perspective as it does not offer clear insight into the uncertainty calibration in accurate and inaccurate regions. Such region-specific insight is useful as high uncertainty in inaccurate regions and low uncertainty in accurate regions can provide heatmaps that could help direct visual attention during QA. Hence, the R-AvU graph uses the building blocks of the AvU metric and plots the uncertainty probabilities in accurate and inaccurate regions across a range of uncertainty thresholds. We use entropy as an uncertainty metric in our experiments, which has been previously shown to capture both data and model uncertainty \cite{uncertainty_gal2016}.

\section{METHOD}
\label{sec:method}

\subsection{Data}
CT scans along with annotations for 9 organs at risk (OAR) in the head-and-neck area were used from the MICCAI 2015-Head and Neck Segmentation Challenge dataset \cite{miccai2015_raudaschl2017}. This dataset provided 33 training and 10 test samples from the RTOG 0522 clinical trial \cite{dataset_rtog_ang2014}. Models trained on this dataset were also evaluated on a separate dataset titled DeepMindTCIA \cite{deepmindtcia_2018} which contains 15 patients. The DeepMindTCIA dataset also refers to the RTOG 0522 clinical trial along with the TCGA-HNSC \cite{dataset_deepmind_tcga_zuley2016} collection on The Cancer Imaging Archive (TCIA). Duplicate RTOG 0522 patients were removed from the DeepMind TCIA dataset if they were already present in the MICCAI dataset. Each CT volume is resampled to a resolution of (0.8, 0.8, 2.5) mm and cropped with a bounding box of dimensions (240,240,80) around the brainstem. The resampling and subsequent training was done at a fixed resolution so that it is convenient for the convolution kernels to learn anatomical feature extraction. The scans were cropped around the brainstem to reduce the computational complexity of patch extraction. The Hounsfield units were trimmed from -125 to +225 to better capture contrast for soft tissues. The models consumed random 3D patches of size (140,140,40) that were augmented with 3D translations, 3D rotations, 3D elastic deformations and Gaussian noise.

\subsection{Neural Architecture}
\label{sec:neuralarch}
The base convolutional neural network (CNN)  of our Bayesian models is inspired by FocusNet \cite{focusnet_gao2019}, a deterministic model. This model is a standard encoder-decoder architecture that uses Squeeze and Excitation \cite{squeeze_hu2018} modules for improved feature extraction via channel attention, a DenseASPP \cite{denseaspp_yang2018} module to obtain sufficient receptive field and finally a supplementary network to prevent foreground-background imbalance for smaller organs at risk (OAR) like optic nerves and optic chiasm. Our implementation avoids the supplementary network for the sake of simplicity. We add Bayesian layers in the DenseASPP module which forms the middle layers of FocusNet. 

A choice of either DropOut \cite{dropout_gal2016} or FlipOut \cite{flipout_wen2018} layers were used for Bayesian modelling. Bayesian modelling of a predictive model involves placing a prior over the models weights $p(W)$ and updating its posterior $p(W|D)$ via observations $D=(X,Y)$ where $X$ and $Y$ are training inputs and outputs respectively. Learning a distribution over the model weights, instead of simply learning fixed scalar values, helps us capture how much the output can vary when provided some input. Thus, Bayesian modelling helps us infer the output distribution $p(y|x,D)$ where $x$ is a test sample and $y$ is its associated output by marginalizing over the posterior:
\begin{equation}
    p(y|x,D) = \mathbbm{E}_{W \sim{p(W|D)}} [p(y|x,W)].
\end{equation}

Theoretically, the DropOut model estimates the posterior distribution of a deep Gaussian process (\textit{a Bayesian inference tool}) by placing a Bernoulli distribution with parameter $p_d$ on the neural net weights. This was shown to be equivalent to performing dropout on the outputs of the layer that those weights belong to. Here output refers to the result of a convolution operation i.e. $w_h * x_h$, where $w_h$ is the kernel weight and $x_h$ is the input in some hidden layer and dropout refers to randomly setting this output to zero with probability $p_d$. FlipOut on the other hand assumes the weight distribution to be Gaussian. In practice, Monte-Carlo sampling via multiple forward passes is used to estimate or infer $p(y|x,D)$. Thus, in every forward pass, Dropout and FlipOut perform output space and weight space perturbations respectively. This is because during each forward pass the DropOut model drops outputs randomly while the FlipOut model samples new weights from a Gaussian distribution. Our DropOut model contains $\sim$500k parameters, while the FlipOut model contains twice those parameters due to the Gaussian assumption. We chose a fixed probability of $p_d=0.25$ for the Dropout model.

\subsection{Training and Inference}
During a single forward pass, the models produce 3D probability maps for each OAR, with each voxel being represented by a vector containing probability values for each OAR that sum to 1. An argmax operator is applied on each voxel's probability vector to assign it an OAR. For each OAR, we assume its 3D predicted probability map to be $P_c$ and the corresponding ground truth probability map to be $Y_c=\{0,1\}$, where $c \in C$ stands for OAR class id. The models are trained using either soft-DICE \cite{vnet_milletari2016} or cross-entropy (CE) loss, which is calculated for each OAR and then averaged to calculate the gradient for back propagation. During training, we perform only a single forward pass to calculate the loss. The DICE loss is calculated as follows:
\begin{align}
    DICE_c &= \frac{2 \displaystyle\sum_{i=1}^{N}(P_{c}^{i} Y_{c}^{i})}{\displaystyle\sum_{i=1}^{N}P_{c}^{i} + \displaystyle\sum_{i=1}^{N}Y_{c}^{i}}, \\
    L_{DICE} &= 1 - \frac{1}{C}\left(w_c\displaystyle\sum_{c=1}^{C}DICE_c\right),
\end{align}
where $P_{c}^{i}$ represents the predicted probability of one of $N$ voxels, $Y_{c}^{i}$ is its corresponding ground truth and $w_c$ is the weight assigned to each class. We use a weighted approach since the OARs in the head and neck region suffer from an imbalanced class problem. The weights are inversely proportional to the average voxel count of each OAR. Similar to DICE, the standard CE loss only penalizes the foreground of each organs probability map i.e. $\mathbbm{1}_{\{Y_c=1\}}$. Our modified CE loss inspired by \cite{modifiedce_taghanaki2019} also penalizes the background i.e. $\mathbbm{1}_{\{(1-Y_c)=1\}}$ of these probability maps for additional supervision as follows:
\begin{align}
    CE_{foreground} &= \displaystyle\sum_{i=1}^{N}(\mathbbm{1}_{\{Y_{c}^{i}=1\}} \ln(P_c^i)) \\
    CE_{background} &= \displaystyle\sum_{i=1}^{N}(\mathbbm{1}_{\{(1-Y_c^i)=1\}} \ln(1 - P_c^i) \\
    L_{CE} &= \frac{1}{C}\left(w_c\displaystyle\sum_{c=1}^{C}(CE_{foreground} + CE_{background})\right),
\end{align}
which showed improved performance when compared to using the standard CE loss.

To train the FlipOut model, one minimizes the CE loss as well as the KL-Divergence term between the Gaussian prior $p(w)$ and the estimated posterior $p(w|D)$ \cite{flipout_wen2018}. For inferring the predictive distribution $p(y|x,D)$ from the model posterior $p(W|D)$, Monte Carlo sampling is performed. We perform $M=30$ forward passes, each time sampling from the posterior to produce 3D activation maps $(P_c)_m$ for each OAR. These are then averaged ($\bar{P_c}$) and passed through the argmax operator to yield the output $\hat{Y}$ containing OAR ids. 
\begin{align}
        \bar{P_c} &= \frac{1}{M} \displaystyle\sum_{m=1}^{M} (P_c)_m \\
        \hat{Y} &= \displaystyle\argmax_{c=1}^{C} \ [\bar{P_c}]
\end{align}

We train and evaluate 4 Bayesian models c.f. DropOut-CE-Basic, DropOut-DICE, DropOut-CE and FlipOut-CE along with some deterministic variants. Here Dropout-CE-Basic is the model trained with the foreground-only cross entropy loss while DropOut-CE is trained with the modified-CE loss described above. In the deterministic (i.e. non-Bayesian) variants c.f. DropOut-DICE-Det and DropOut-CE-Det, only a single forward pass (i.e $M=1$) is performed. A deterministic analysis on FlipOut-CE is not done as its design leads to new weights being sampled in every forward pass. The models were trained for a 1000 epochs with the Adam optimizer and a fixed learning rate of 0.001, with one epoch looping over 33 patients in the MICCAI2015 training subset. 

\subsection{Uncertainty}
Using the probability maps $(P_c)_m$ of each OAR, we compute the  entropy maps and use them as uncertainty maps. Entropy is a term derived from information theory that captures the average amount of uncertainty present in a signal. Thus, if Monte Carlo ($M$) sampling in a Bayesian network leads to highly varying probability vectors for a voxel, it would have higher entropy. To calculate the 3D entropy map $H(y|x,D)$, we use the averaged probability heatmaps $\bar{P_c}$ of each OAR:
\begin{equation}
    \begin{aligned}
        H(y|x,D) = -\displaystyle\sum_{i=1}^{C} \bar{P_c}\cdot\log(\bar{P_c}),
    \end{aligned}
\end{equation}
which has a maximum value when the average probability vector $\bar{P_c^i}$ for each voxel $i$ has all its values as $\frac{1}{C}$. In our case of C=10 (9 OARs + background), the maximum entropy value is 2.3.   



\begin{figure}[tb]
    \centering
    \begin{subfigure}
      \centering
      \includegraphics[height=5.0cm, trim=0 0 340 0, clip]{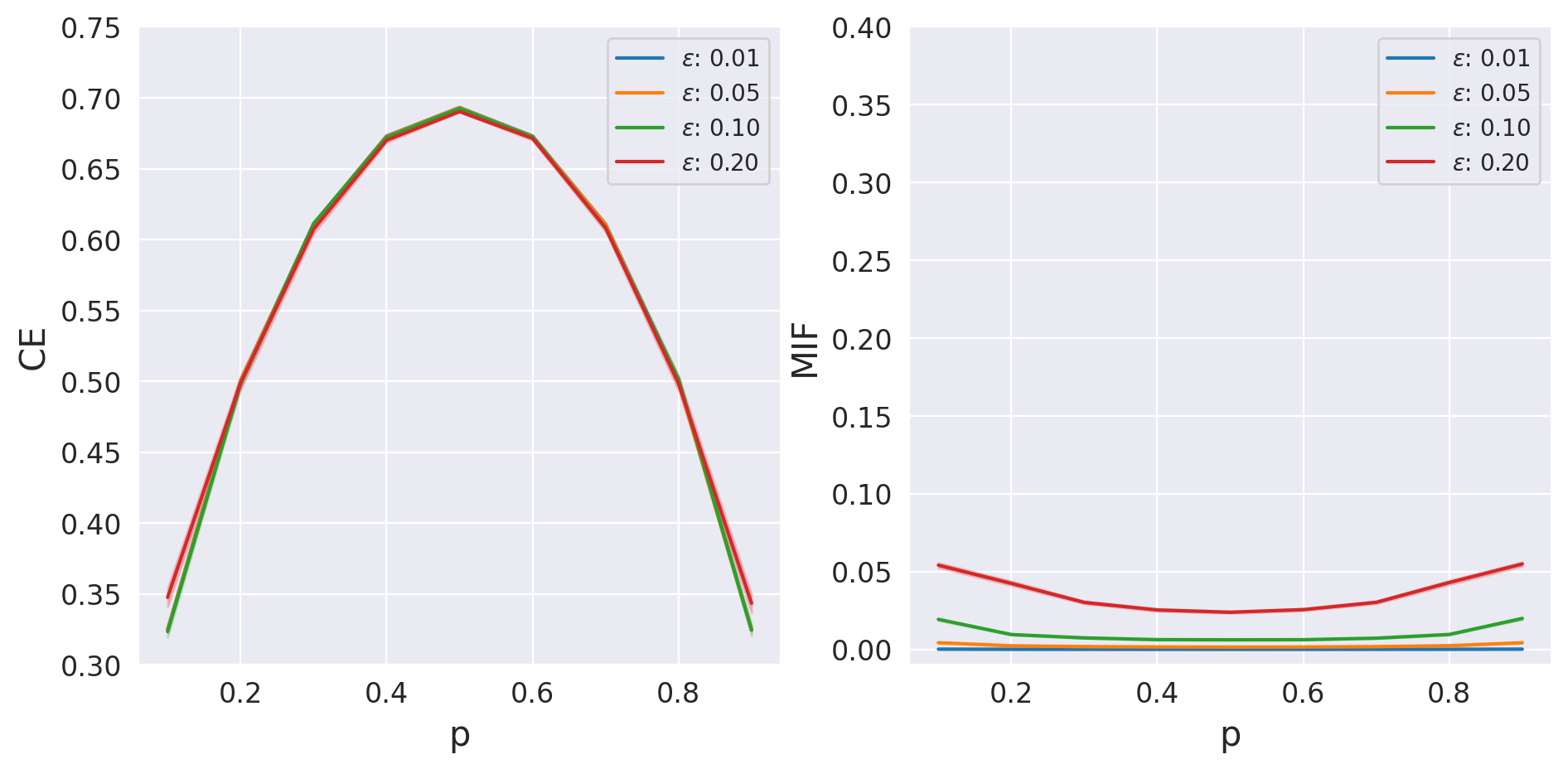}
    \end{subfigure}%
    \begin{subfigure}
      \centering
      \includegraphics[height=5.0cm, trim=0 0 340 0, clip]{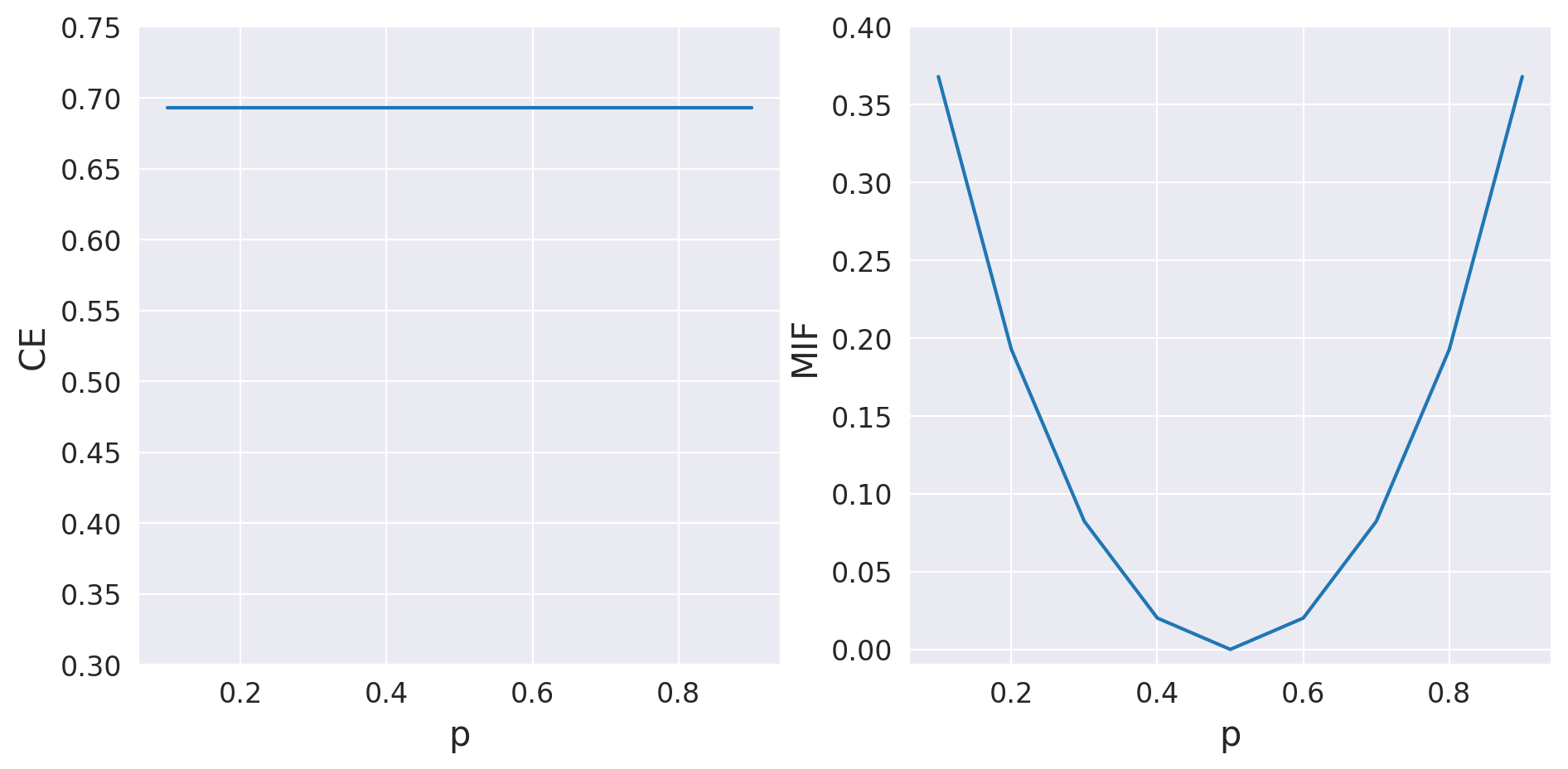}
    \end{subfigure}
    \caption{These figures show the behaviour of entropy for a simple binary classification problem of one voxel. Here $p$ represents the foreground class probability and $\epsilon$ refers to the amount of output probability variability across Monte Carlo runs. The left figure shows uncertainty behavior when the output probability has some variability, while the right figure shows uncertainty behaviour in case of extreme probability changes.}
    \label{fig:fig_unc_behaviour}
\end{figure}

In Figure \ref{fig:fig_unc_behaviour}, we use a toy binary classification problem (e.g. foreground vs background classification for a single voxel) to understand the behavior of these metrics. In the left figure, we add uniform variability parameterized by $\epsilon$ to the foreground class probability $p$ to replicate possible Monte Carlo outputs. Here, entropy is maximum at $p = 0.5$, i.e the model assigns equal probability to both foreground and background. It is lowest when the model is confident in its predictions i.e. $p = \{0, 1\}$. Also, while increasing the amount of variation across different Monte Carlo outputs, there is no behavioral change in entropy as seen by the overlap of the curves. In the right figure, we investigate an extreme case wherein Monte Carlo sampling outputs probabilities such as $[p, 1-p, p, \ldots]$. This replicates extreme probability swings which might represent the case of a boundary voxel between an OAR and background where contrast is poor and hence the model is uncertain. Such outputs maximize the entropy across all probabilities.

\subsection{Evaluation}
For evaluation, we use two metrics: the expected calibration error (ECE) \cite{ece_guo2017} for model confidence calibration and then region-based accuracy-vs-uncertainty (R-AvU) for uncertainty calibration. For e.g. in a foreground-background classification problem, if 100 voxels are assigned the foreground class with 70$\%$ probability, then we should expect that 70 of those voxels have been assigned the correct class. The error between the model confidence and its accuracy is considered as calibration error. When the same is averaged across multiple probability bins, we obtain the expected calibration error. Specifically, for each OAR, we calculate $ECE_c$ by assigning the probability of each predicted OAR voxel $i$ to one of B=10 equally spaced bins ($B_p)$ between 0 and 1 as follows:
\begin{align}
        \mathrm{ECE}_c &= \frac{1}{B} (\mathrm{acc}(B_p) - \mathrm{conf}(B_p)), \\
        \mathrm{acc}(B_p) &= \frac{1}{|B_p|} \displaystyle\sum_{i \in B_p} \mathbbm{1}_{\hat{Y_c} = Y_c}, \\
        \mathrm{conf}(B_p) &= \frac{1}{|B_p|} \displaystyle\sum_{i \in B_p} (P_c)_i.
\end{align}
Here $Y_c$ is the ground truth map, $\hat{Y_c}$ is the predicted map and $P_c$ is the probability map belonging to a particular OAR. The lower the ECE values, the more calibrated a model is. Finally, to compute the R-AvU graphs we use uncertainty heatmaps to create line plots for the probability of uncertainty in inaccurate ($p(u|i)$) regions as well as the probability of uncertainty in accurate ($p(u|a,\sim{a})$) regions. In the context of this graph, each voxel has two properties: its accuracy and uncertainty. Each voxel is then categorized as $n_{ac}$, $n_{au}$, $n_{ic}$ and $n_{iu}$ where $n$ stands for the number of voxels, $a$ for accurate, $i$ for inaccurate, $c$ for certain and $u$ for uncertain. Using these terms, we find the two curves in the R-AvU graph
\begin{align}
        p(u|i) &= \frac{n_{ui}}{n_{iu} + n_{ic}} \\
        p(u|a,\sim{a}) &= \frac{n_{au}}{n_{au} + n_{ac}}
\end{align}

We define accurate regions as those containing true positive (TP) voxels. We include the $\sim{~}a$ term to denote \textit{almost} TP voxels, as due to inter- and intra-observer variation, it is common to disregard false positive (FP) and false negative (FN) voxels very close to the ground truth contours. This is done by an erosion followed by a dilation on the inaccurate regions using a (3,3,1) filter which removes any small regions of error. The remaining FP and FN voxels are then considered as the inaccurate regions. Such an interpretation may be useful for radiotherapy QA, where smaller contouring errors may not have significant downstream effects on the calculated radiation dose for healthy tissue. Thus, such areas can be considered accurate enough and it is preferable from a visual attention standpoint that a model has lower uncertainty in these regions.


\section{RESULTS}
\subsection{Volumetric Performance} 
 
\begin{figure}[tb]
    \centering
    \begin{subfigure}
      \centering
      \includegraphics[height=4.0cm]{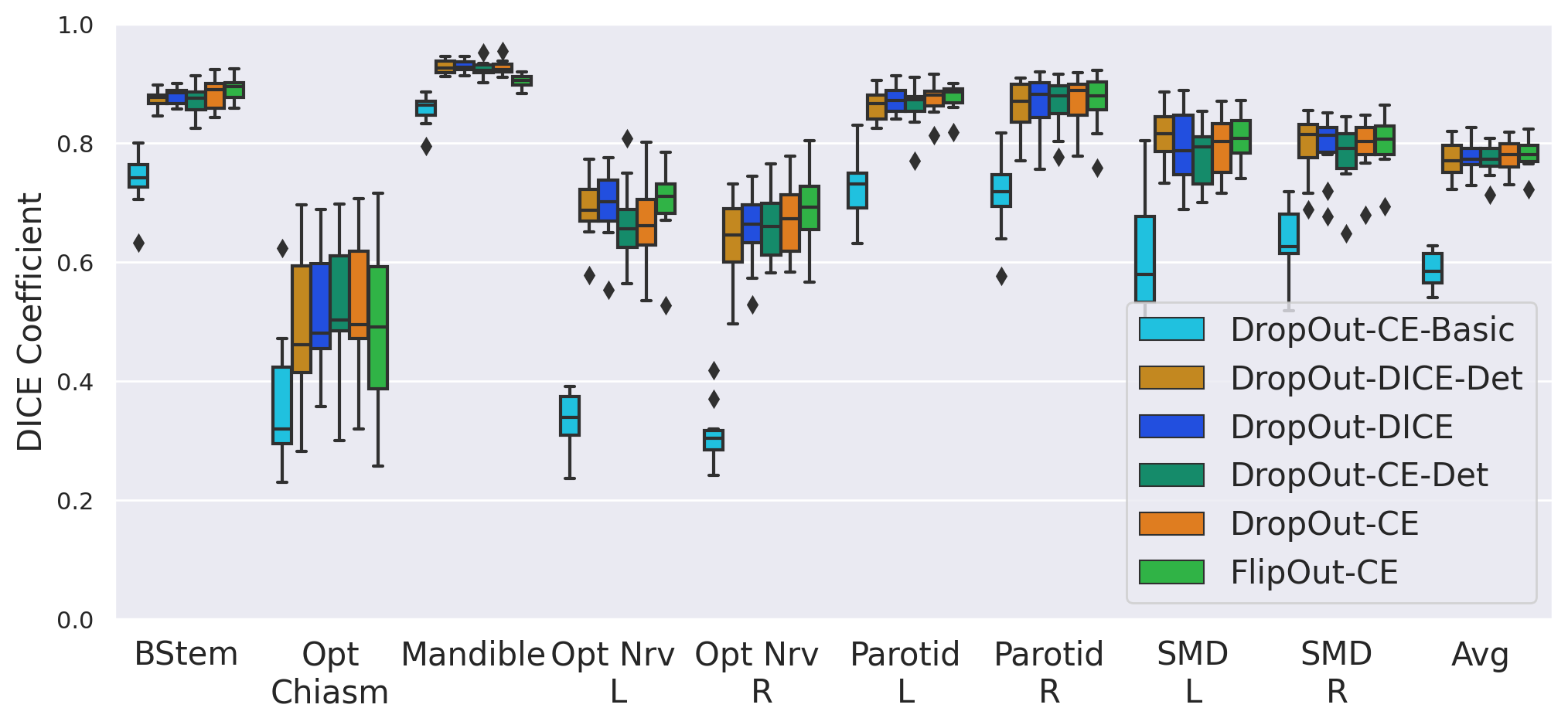}
    \end{subfigure}%
    \begin{subfigure}
      \centering
      \includegraphics[height=4.0cm]{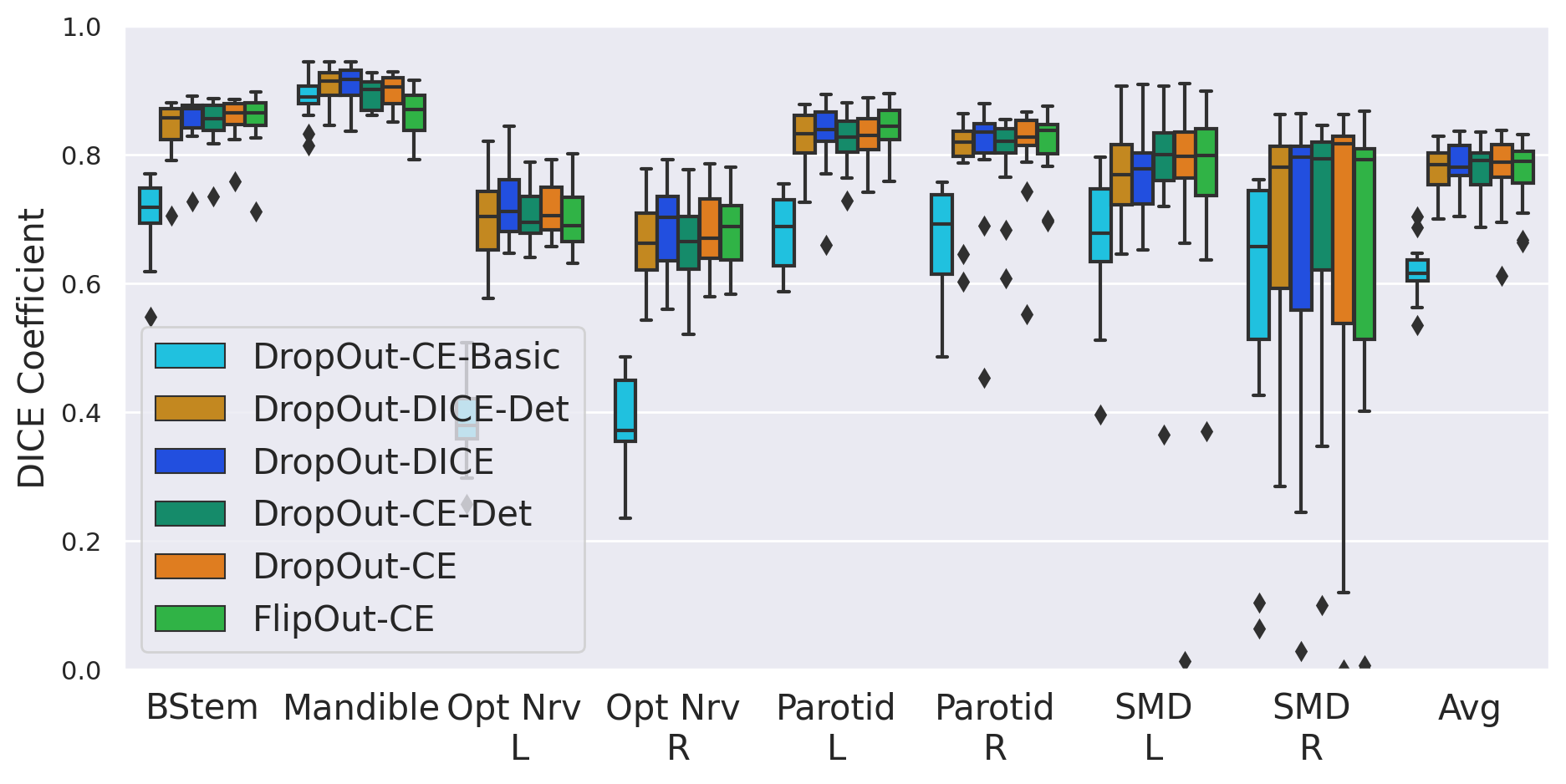}
    \end{subfigure}
    \caption{Boxplot depicting the DICE scores for the MICCAI2015 test dataset (left) and the DeepMindTCIA dataset (right). The x-axis shows the different organs and the average over all organs.}
    \label{fig:fig_dice}
\end{figure}

Figure \ref{fig:fig_dice} shows OAR DICE scores for the MICCAI 2015 test dataset on the left and for the DeepMindTCIA dataset on the right. For both datasets, the mandible and the brainstem (BStem) achieve the highest scores followed closely by the parotid and submandibular (SMD) glands while the optic organs (Opt Nrv L, Opt Nrv R and Opt Chiasm) have lower DICE scores overall. In the DeepMindTCIA dataset, we see various outliers for the right submandibular gland (SMD R). For the MICCAI 2015 test dataset, all models, except DropOut-CE-Basic have equivalent average performance in terms of standard medical segmentation metrics, i.e DICE ($\sim$0.77 - 0.78) and Hausdorff Distance 95$\%$ ($\sim$5mm - 7mm). We run a Wilcoxon signed-rank test on the Bayesian models and achieve p-values of 0.625 between DropOut-DICE and DropOut-CE, 0.275 between DropOut-DICE and FlipOut-CE and 1.0 between DropOut-CE and FlipOut-CE for the average DICE scores. For the average Hausdorff Distance 95$\%$ we achieve p-values of 0.027 between DropOut-DICE and DropOut-CE, 0.375 between DropOut-DICE and FlipOut-CE and 0.232 between DropOut-CE and FlipOut-CE. The results indicate that for the most part the models are not significantly different. Thus, we may compare these models using other metrics such as Expected Calibration Error (ECE) and Region-Accuracy vs Uncertainty (R-AvU). No statistical tests or additional metrics were used to study the DropOut-CE-Basic model due to its poor performance on average DICE (0.58) and average Hausdorff Distance 95$\%$ (15.95mm). Tensorflow \cite{tensorflow2015-whitepaper} code to reproduce these results can be found at \href{https://github.com/prerakmody/hansegmentation-uncertainty-qa}{https://github.com/prerakmody/hansegmentation-uncertainty-qa}.

\subsection{Expected Calibration Error}
\begin{figure}[tb]
    \centering
    \begin{subfigure}
      \centering
      \includegraphics[height=4.0cm]{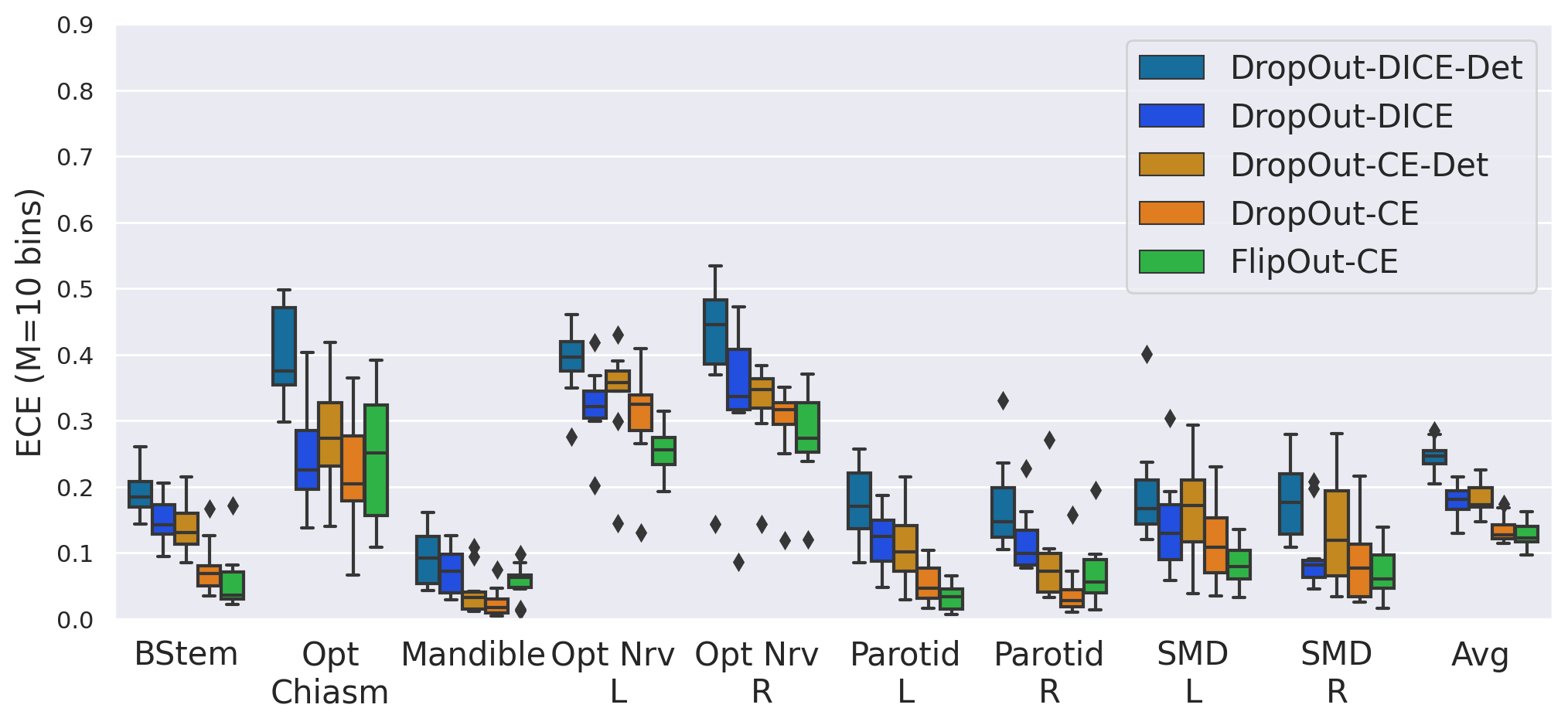}
    \end{subfigure}%
    \begin{subfigure}
      \centering
      \includegraphics[height=4.0cm]{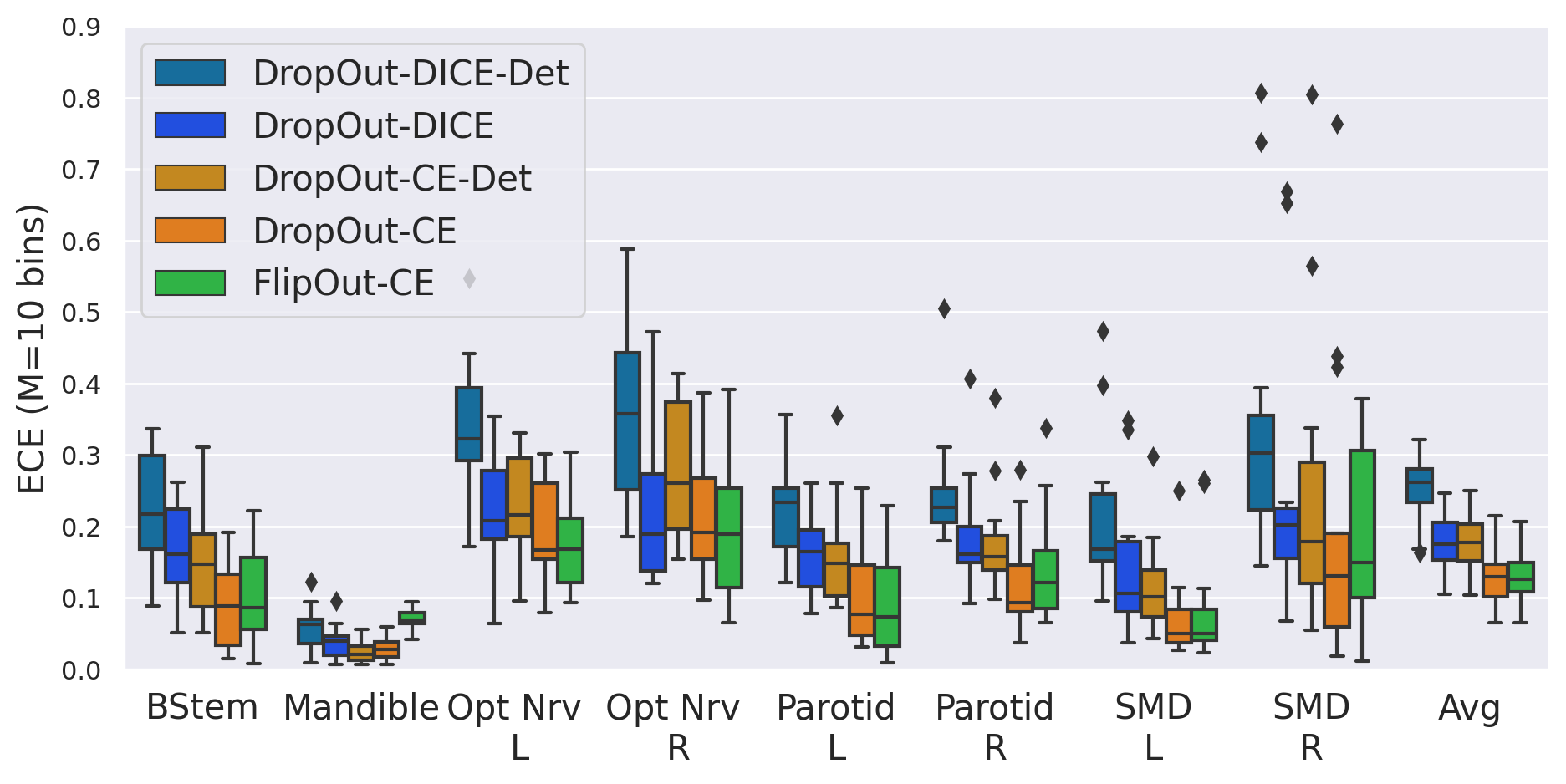}
    \end{subfigure}
    \caption{Boxplot depicting the Expected Calibration Error (ECE) with M=10 bins for the MICCAI2015 test dataset (left) and the DeepMindTCIA dataset (right). The x-axis shows the different organs and the average.}
    \label{fig:fig_ece}
\end{figure}

Figure \ref{fig:fig_ece} shows for both datasets that Dropout-DICE and DropOut-CE always have lower ECE than their deterministic counterparts Dropout-Dice-Det and DropOut-CE-Det. DropOut-CE on average has a lower ECE than DropOut-DICE, while FlipOut-Det and FlipOut-CE have similar ECE. The same holds for DropOut-CE and FlipOut-CE. For organs, we notice that the optic organs have the highest ECE compared to other organs for both datasets. The submandibular glands (SMD L and SMD R) and the right parotid gland have outliers in the DeepMindTCIA dataset as shown on the right side of Figure \ref{fig:fig_ece}.  

\subsection{Region - Accuracy vs Uncertainty}
\begin{figure}[tb]
    \centering
    \begin{subfigure}
      \centering
      \includegraphics[height=4cm]{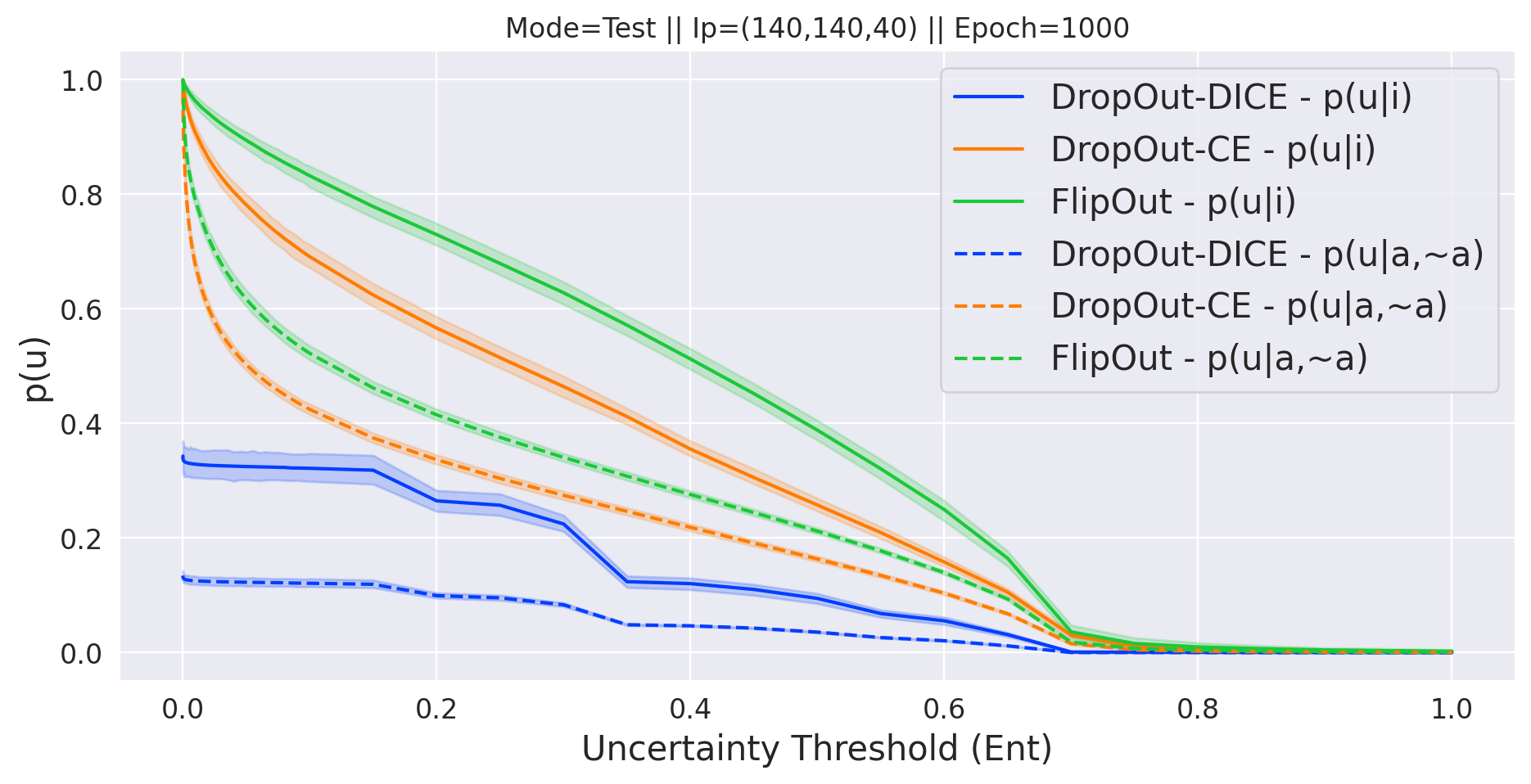}
    \end{subfigure}%
    \begin{subfigure}
      \centering
      \includegraphics[height=4cm]{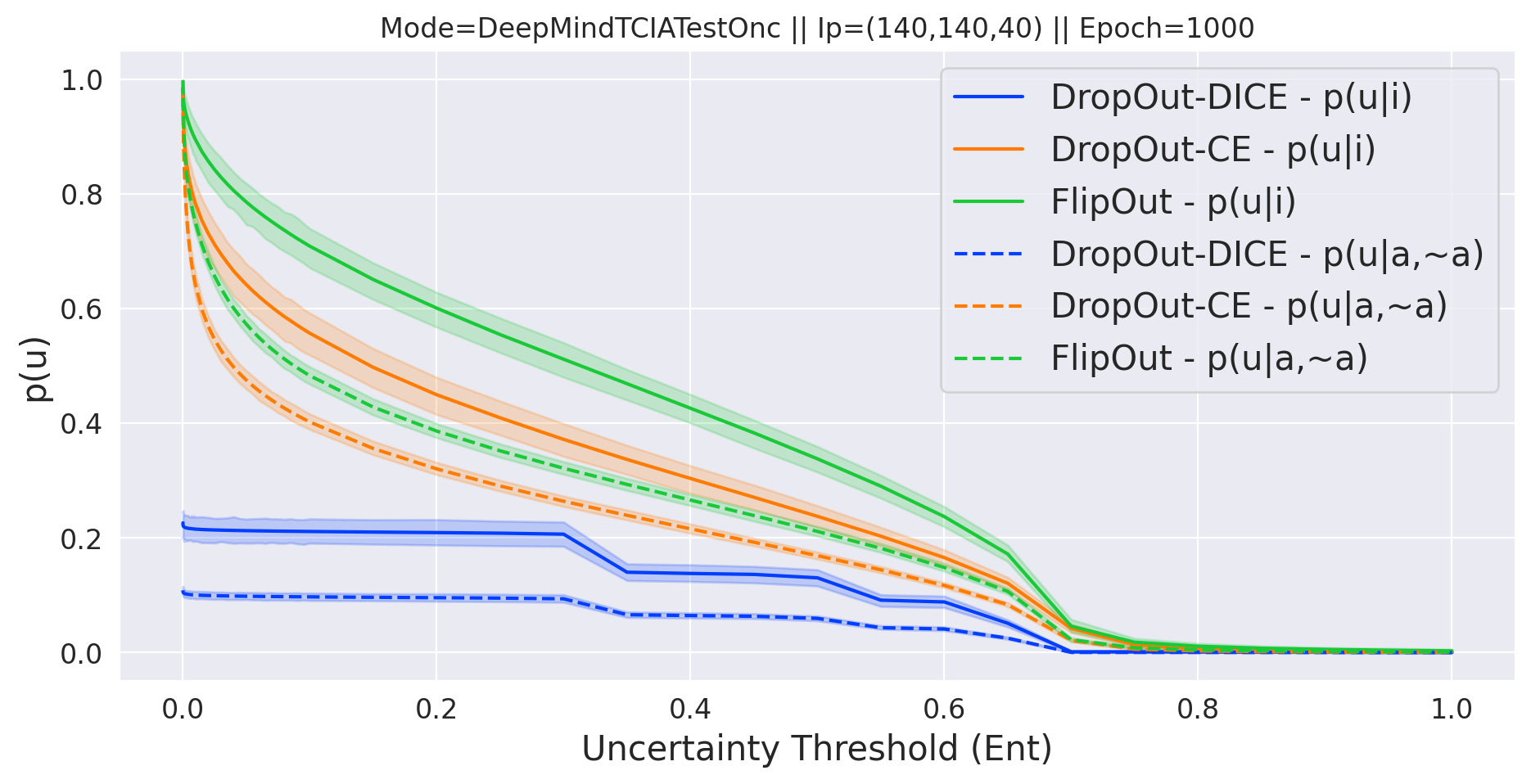}
    \end{subfigure}
    \caption{Line plots showing the uncertainty behaviour of different models in inaccurate ($p(u|i)$) and accurate ($p(u|a,\sim{a})$) regions for the MICCAI2015 test set (left) and the DeepMindTCIA dataset (right).}
    \label{fig:fig_pu_ent}
\end{figure}

Figure \ref{fig:fig_pu_ent} represents $p(u|i)$ as a solid line plot and  $p(u|a,\sim{a})$ as a dotted line plot for entropy. A model for efficient QA would have high $p(u|i)$ and low $p(u|a,\sim{a})$. The $p(u|i)$ and $p(u|a,\sim{a})$ of the FlipOut-CE model is higher than that of the DropOut-CE model for the entire range of uncertainty thresholds. For entropy as the uncertainty metric, the DropOut-DICE model always has values lower than DropOut-CE and FlipOut-CE for both $p(u|i)$ and $p(u|a,\sim{a})$. Similar trends are noticed for the DeepMindTCIA dataset, though the probability values are slightly reduced.

\begin{figure}
    \centering 
    \includegraphics[height=7.3cm]{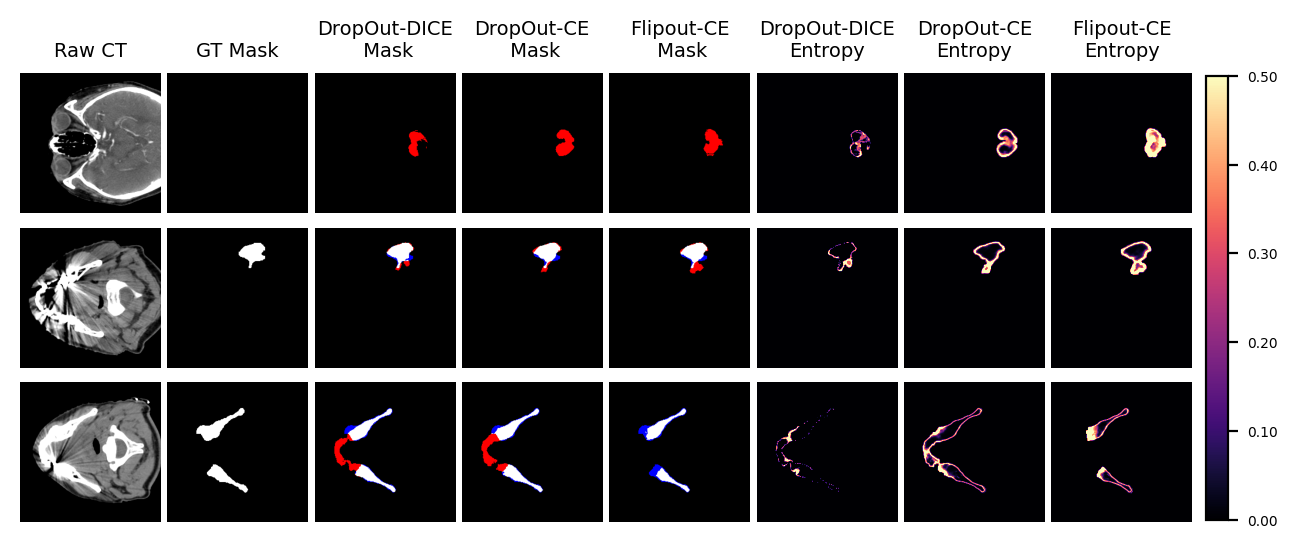}
    \caption{The first two columns depict the raw and ground truth data from the datasets, while the remaining columns show model predictions and their associated entropy heatmaps. In the predicted masks, white voxels are true positives, red voxels are false positives while blue voxels are false negatives.}
    \label{fig:fig_examples_ent}
\end{figure}

For visual results, we look at Figure \ref{fig:fig_examples_ent} where the first column shows a CT slice and the second column shows the ground truth (GT) mask. The third, fourth and fifth columns are the deep learning predictions and the remaining columns are their corresponding uncertainty heatmaps. The first row in the figure shows a result from the MICCAI2015 test dataset representing a false positive prediction for the top slice of the brainstem. The second and third rows show predictions for the DeepMindTCIA dataset of the left parotid gland and mandible respectively. In these figures, red represents false positive, blue represents false negative and white represents true positive predictions.

\section{DISCUSSION AND CONCLUSION}
This work exploited an existing deterministic model (i.e. FocusNet \cite{focusnet_gao2019}) and investigated the model confidence calibration and uncertainty behavior of its Bayesian versions for efficient QA in a clinical radiotherapy setting. All Bayesian models, when averaged across organs at risk (OAR), performed equally well in terms of volumetric and surface distance measures, allowing us to compare across other metrics like expected calibration error (ECE) and region-based accuracy-vs-uncertainty (R-AvU). Using a modified cross entropy loss for our models improved their performance in comparison to its standard version as additional supervision is provided for both the foreground and background of each OAR. It was also important to use weights for each OAR to handle the problem of class imbalance. The right plot in Figure \ref{fig:fig_dice} shows low DICE scores for the right submandibular gland (SMD R) in the DeepMindTCIA dataset. This is because, in general, our models have reduced performance for the TCGA-HNSC patients when compared to the RTOG 0522 patients due to poor contrast in the TCGA-HNSC CT scans.


Post model training, it is important to evaluate the ECE of a predictive model to check if it produces probability estimates that reflects its true underlying interpretation of a test sample. The boxplots in Figure \ref{fig:fig_ece} shows that performing Bayesian inference in neural networks always reduces or maintains calibration error (ECE). Thus, all subsequent model comparisons in this work only consider Bayesian models. It is also observed that CE as a loss function leads to reduced ECE compared to DICE, as also found by others\cite{calibrated_sander2019}. This may be since CE is a strict scoring rule and hence achieves more reliable probability estimates. Also note that the modified CE achieved similar accuracy compared to DICE. This is an important result as most works in medical image segmentation rely on using the DICE loss. Once again, similar to DICE performance, the right submandibular gland (SMD R) in the DeepMindTCIA dataset has outlier ECE values. This is due to the fact the models are highly confident but yet inaccurate, leading to large calibration errors.

Given that DropOut-CE and FlipOut-CE have similar ECE values, we refer to the R-AvU graphs to understand differences in their behavior in the context of output uncertainty. For entropy, the FlipOut-CE model has better uncertainty coverage than other models in inaccurate regions. This is reflected in Figure \ref{fig:fig_pu_ent} where both its $p(u|i)$ and $p(u|a,\sim{a})$ curves are higher than that of DropOut-CE. This means that FlipOut-CE misses less inaccurate regions than DropOut-CE, but also directs visual attention to areas that are accurate, more so than DropOut-CE, potentially slowing down QA. A possible reason for the behavior of FlipOut-CE could be that it uses a Gaussian distribution which might be more representative of the weight distribution than the Bernoulli distribution. Entropy for Dropout-DICE, which has the highest ECE, has uncertainty curves that do not sufficiently cover incorrect regions, thus reducing its potential as a contour QA candidate. 

Focusing on the bright areas in Figure \ref{fig:fig_examples_ent}, the first and third row show that FlipOut-CE provides a better coverage of erroneous regions, while in the second row the bright areas of DropOut-DICE correspond to errors in the different lobes of the left parotid gland. In the third row of Figure \ref{fig:fig_examples_ent} for CE-trained models, we see that there exists high uncertainty in the erroneous regions and low uncertainty along the borders of the mandible. The low uncertainty could be the effect of different annotation quality for different patients  in the training data which leads to data-based uncertainty along the border regions of an OAR. A similar effect for CE-trained models is seen in row 2 for the left parotid gland, but in this case there is high uncertainty in both high and low error regions which does not satisfy our requirements for visual attention. It is due to this effect that the $p(u|a)$ curves have high probability values. Finally, uncertainty does not exactly correspond to voxel-wise error, so an additional visualization tool on top of the output uncertainty heatmaps may improve acceptability from clinical users. 

To conclude, we show that considering both foreground and background regions in the probability maps of organs for the cross entropy (CE) loss improves model performance over the standard practice of only using the foreground regions. This is beneficial, as CE-trained models have better model confidence calibration than DICE trained models. We also explored how the combined use of a quantitative and qualitative measure can support the analysis and selection of Bayesian models for radiotherapy QA. It was observed, that on average, FlipOut-CE has more uncertainty coverage of both inaccurate and accurate regions than the DropOut models, possibly due to the Gaussian assumption in FlipOut compared to the Bernoulli assumption in DropOut. Future work may consider additional training objectives to push apart the $p(u|i)$ and $p(u|a,\sim{a})$ curves with the $p(u|i)$ curve having high values and the $p(u|a,\sim{a})$ curves having lower values. This will ensure visual attention in erroneous regions through the use of uncertainty heatmaps. One may also explore the use of uncertainty metrics like mutual information that only capture model uncertainty \cite{uncertainty_gal2016}, unlike entropy that captures both data and model uncertainty. It might be worthwhile to investigate which uncertainty metric is more useful within clinical workflows. Finally, this study could also be done for a contour propagation scenario in adaptive radiotherapy to observe if similar results are obtained.

\acknowledgments 
The research for this work was funded by the HollandPTC-Varian Consortium (grant id 2019022).

\bibliography{report} 
\bibliographystyle{spiebib} 

\end{document}